\documentclass[11pt,a4paper,english,preprint,aps,nofootinbib]{revtex4}
 \pdfoutput=1
\usepackage[english]{babel}
\usepackage{graphicx,color}
\usepackage{latexsym}
\usepackage{amsmath}
\usepackage{amsfonts}
\usepackage{amssymb,slashed}
\usepackage{bbold}
\usepackage[latin1]{inputenc}

\begin{document}

\title{Novel Studies on the $\eta'$ Effective Lagrangian}
\author{\textbf{A. P. Balachandran}\footnote{bal@phy.syr.edu} $^{a,b}$}

\author{\textbf{T. R. Govindarajan}\footnote{trg@imsc.res.in} $^{b}$}

\author{\textbf{Amilcar R. de Queiroz}\footnote{amilcarq@unb.br} $^{b,c}$}  

\affiliation{$^a$ Physics Department, Syracuse University, Syracuse, NY, 13244-1130, USA}

\vspace*{1cm}

\affiliation{$^b$ Institute of Mathematical Sciences, CIT Campus, Taramani, Chennai 600113, India}

\vspace*{1cm}

\affiliation{$^c$ Instituto de Fisica, Universidade de
Brasilia, Caixa Postal 04455, 70919-970, Brasilia, DF, Brazil}

\preprint{IMSc/2012/2/1}

\begin{abstract}
The effective Lagrangian for $\eta'$ incorporating the effect of the QCD $\theta$-angle has been developed previously. We revisit this Lagrangian and carry out its canonical quantization with particular attention to the test function spaces of constraints and the topology of the $\eta'$-field. In this way, we discover a new chirally symmetric coupling of this field to chiral multiplets which involves in particular fermions. This coupling violates $P$ and $T$ symmetries. In a subsequent paper, we will evaluate its contribution to the electric dipole moment (EDM) of fermions.  Our motivation is to test whether the use of mixed states restores $P$ and $T$ invariance, so that EDM vanishes. This calculation will be shown to have striking new physical consequences. 
\end{abstract}
\maketitle

\tableofcontents

\section{Introduction}

The $\eta'$ meson obtains its mass from the axial anomaly in QCD. The origin of this mass is not just the local form of the anomaly, but also non-perturbative effects coming from instantons.

The QCD $\theta$-angle also originates from non-perturbative instantons effects. Its source is the multiple connectivity of the configuration space in QCD \cite{balachandran1991classical} and does not require the presence of spin $1/2$ fields and the attendant axial anomaly.

But there is all the same a deep relation between axial anomaly and QCD $\theta$. The Peccei-Quinn \cite{PhysRevLett.38.1440,PhysRevD.16.1791} mechanism, basic to discussions of the CKM matrix \cite{PTP.49.652}, is an example of this connection.

Earlier work by Rosenzweig \emph{et al.} \cite{PhysRevD.21.3388}, Aurilia \emph{et al.} \cite{Aurilia1980} and others \cite{Veneziano:1979ec,Crewther:1979pi,ohta-1980, ohtaa-1981,ohtab-1981,PospelovAnnalsPhys.318:119-1692005,Dar2009}, developing the theory of 't Hooft \cite{Hooft1986}, has formulated the effective Lagrangian for $\eta'$ in the presence of instantons. In this and subsequent papers, we reexamine this model from the canonical point of view. The analysis involves important new considerations on the test function spaces of first class constraints which have not been treated in the literature. In particular, the QCD $\theta$-states constructed using an analogue of the Chern-Simons term is ``gauge invariant" or is annihilated by a Gauss law constraint only after the proper choice of the test function space for constraints is identified. 

Our research was motivated by an attempt to revisit the calculation of electric dipole moment (EDM) from the effective Lagrangian approach and to test whether the use of mixed states \cite{Balachandran2012,Balachandran:2011gj} restores $P$ and $T$ invariance, so that the EDM becomes zero. As a prelude to this calculation, to be reported in a following paper \cite{Bal-edm-2012}, here we formulate the chiral invariant coupling of $\eta'$ to nucleons or quarks which is sensitive to the QCD $\theta$. It improves upon the standard treatment of this coupling. The fermion mass term is affected if $\sin \theta \neq 0$. We will show in a coming paper that it induces EDM \cite{Bal-edm-2012}. We have however pointed out earlier \cite{Balachandran2012} that the use of mixed states makes the expectation value of EDM zero.

In section II, we recall the effective Lagrangian for $\eta'$ and the demonstration that $\eta'$ has mass because of the axial anomaly. In section III, we develop the Hamiltonian formalism and constraint analysis with particular attention to boundary conditions on test functions. The QCD $\theta$-states are then formulated after making essential use of the properties of test function spaces. They are ``gauge invariant" and hence are annihilated by the Gauss law constraints. In section IV, we present a brief, but not quite obvious, demonstration, that the Hamiltonian approach also displays a mass gap and describes a massive $\eta'$. In section V, we formulate the chirally invariant couplings of quarks and nucleons to the chiral multiplet containing $\eta'$ as well. We also show that the fermion mass term has a $\sin \theta$ dependent $P$ and $T$ violating piece. It modifies the fermion propagator. The latter will be the basis of the electric dipole moment calculation in a following paper.

\section{The $\eta'$ Model}

The target space for the chiral model of $N_f$ flavors is the group manifold of $U(N_f)$. The chiral group $SU(N_f)_L\times SU(N_f)_R$ acts on $U(N_f)$ according to the following rule:
      \begin{align}
            (g_L,g_R) \triangleright u(x)~\rightarrow~ g_L u(x) g_R^\dagger, 
      \end{align}
for $g_L\in SU(N_f)_L$, $g_R\in SU(N_f)_R$ and $u(x)\in U(N_f)$. The action of the axial vector group $U(1)_A$ instead is
\begin{equation}
      u(x)\rightarrow \omega u(x), ~~~~ |\omega|=1.
\end{equation}

\subsection{The Topology of the Target Space}

The topology of the target manifold $U(N_f)$ is important for us. We have that 
\begin{align}
      U(N_f) &= \frac{SU(N_f)\times U(1)}{\mathbb{Z}_{N_f}} \\
      \mathbb{Z}_{N_f} &=  \left\langle \left(e^{ik\frac{2\pi}{N_f}} \mathbb{1}_{N_f\times N_f}, e^{-ik \frac{2\pi}{N_f}} \right): k=0,1,...,N_f-1 \right\rangle. 
\end{align}
That is, $U(N_f)$ is not $SU(N_f)\times U(1)$.

In practice what this means is the following: If $s(x)\in SU(N_f)$ describes a flavor multiplet which has pions, and $e^{i\eta'(x)}\in U(1)$ describes the $\eta'$ field, then
\begin{equation}
      \label{u-field-1}
      u(x)=s(x)e^{i\eta'(x)}
\end{equation}
so that the Lagrangian or Hamiltonian should be invariant under the substitutions
\begin{align}
      \label{field-transf-1}
      s(x)\rightarrow s(x) e^{ik\frac{2\pi}{N_f}}, \\
      e^{i\eta'(x)}\rightarrow e^{i\eta'(x)} e^{-i k \frac{2\pi}{N_f}}, \label{field-transf-2}
\end{align}
or
\begin{equation}
      \label{field-transf-3}
      \eta'(x)\rightarrow \eta'(x) - k\frac{2\pi}{N_f}.
\end{equation}

The above simple remark plays a crucial role in what follows.

\subsection{The Lagrangian}

We next focus on the Lagrangian density $\mathcal{L}$ describing just $\eta'$. It is invariant under Eqs. (\ref{field-transf-1},\ref{field-transf-2}). We have
\begin{equation}
      \label{Lagran-eta-prime-1}
      \mathcal{L}=-\frac{1}{2R^2} \left(\partial_\mu \eta'\right) \left(\partial^\mu \eta'\right)+ \lambda  \left(\partial_\mu \eta'\right) B^\mu + \theta \lambda \partial_\mu B^\mu + \frac{1}{2} \left(\partial_\mu B^\mu  \right)^2,
\end{equation}
with $\lambda\in\mathbb{R}\diagdown \{ 0 \}$. We use the metric $(+,+,+,-)$. 

The field $B^\mu$ is dual to the $3$-form $A_{\alpha\beta\gamma}$ used in the literature:
\begin{equation}
      B^\mu=\frac{1}{3!} \epsilon^{\mu \alpha \beta \gamma} A_{\alpha\beta\gamma}.
\end{equation}
Notation is simplified by using $B$ instead of its dual.

\subsection{Equations of Motion}

Here we recall the calculation showing that Eq. (\ref{Lagran-eta-prime-1}) describes a massive field. We obtain from the variations of the action $S=\int d^4x\mathcal{L}$,
\begin{align}
      \delta_{\eta'} S = 0 &\Rightarrow ~~  \frac{1}{R^2} \partial^2 \eta' - \lambda \partial\cdot  B =0, \\
      \delta_{B} S = 0 &\Rightarrow ~~ \lambda \partial_\mu \eta' - \partial_{\mu}\left(\partial\cdot B \right) = 0.
\end{align}
So,
\begin{align}
      \partial^2 (\partial\cdot B) - (\lambda R)^2 \partial\cdot B =0.
\end{align}
Therefore $\partial\cdot B$ is a field of mass $|\lambda R|$. Let us assume that $\lambda>0$, if necessary by changing $B^\mu$ to $-B^\mu$, so that the mass is just $\lambda R$.

\section{Canonical Quantization}

We define the momentum fields conjugate to $\eta'$ and $B^\mu$ by $\pi$ and $P_\mu$, respectively. Then from $\mathcal{L}$, we obtain
\begin{align}
      \pi &= \frac{1}{R^2}\partial_0 \eta' + \lambda B^0, \label{momentum-eta'}\\
      P_0 &= \partial\cdot B + \theta\lambda, \label{momentum-B0}\\
      P_i &\approx 0 ~~~ (\textrm{the primary constraint}) \label{constraint-1}, 
\end{align}
where the symbol $\approx$ stands for weak equality as usual.

The Hamiltonian density is given by
\begin{align}
      \label{hamiltonian-eta}
      \mathcal{H} &= \frac{R^2}{2}\left(\pi - \lambda B^0 \right)^2 + \frac{1}{2} \left( P_0 - \theta \lambda\right)^2 + \frac{1}{2R^2}\left( \partial_i \eta'\right)^2 \nonumber \\
      &- \left(\partial_i B^i \right)\left(P_0-\lambda\eta' \right) + v^i P_i,      
\end{align}
with $v^i$ being Lagrangian multipliers. The Hamiltonian is given by
\begin{equation}
      \label{hamiltonian-eta-2}
      H=\int d^3x~\mathcal{H}(x).
\end{equation}

The secondary constraint follows from $\{ P_i(\mathbf{x},t),\mathcal{H}(\mathbf{y},t)\}\approx 0$, where $\{\cdot,\cdot\}$ denotes the Poisson bracket. It is thus
\begin{equation}
      \label{constraint-2}
      \partial_i\left( P_0 - \lambda \eta'\right) \approx 0.
\end{equation}
There is no tertiary constraint.

The constraints Eq. (\ref{constraint-1}) and Eq. (\ref{constraint-2}) are first class.

\subsection{On Test Function Spaces for Constraints}

Let us first make a motivating remark.

Observe that the following term in the Hamiltonian Eq. (\ref{hamiltonian-eta-2}),
\begin{equation}
      \label{gauss-law-in-Hamilt-1}
      -\int d^3 x~(\partial_i B^i)(x)~(P_0-\lambda\eta')(x)
\end{equation}
is \emph{not} invariant under the transformations Eq. (\ref{field-transf-1}) unless
\begin{equation}
      \label{constraint-condition-1}
      \lim_{r\to \infty} \int d\Omega~r^2 \hat{x}_i B^i (x) = 0,
\end{equation}
that is, this boundary term vanishes at infinity.

We now argue that this is precisely what we need to make this term a constraint.

On quantization, classical fields become quantum fields which are operator-valued \emph{distributions}. But derivatives of distributions have to be understood and expressed in terms of their test functions which make up their dual space. Thus if $c^i$ are test functions for the constraints Eq. (\ref{constraint-2}), we must interpret its quantum version as the equation
\begin{equation}
      \label{quantum-constraint-2}
      \int d^3 x ~\left( \partial_i c^i\right) \left( P_0 - \lambda \eta'\right) (x) |\cdot\rangle = 0
\end{equation}
which constrains the state vectors $|\cdot\rangle$ in the domain of the Hamiltonian.

But in the classical limit, Eq. (\ref{quantum-constraint-2}) must go over to Eq. (\ref{constraint-2}). Classically, 
\begin{align}
      \label{constraint-condition-2}
      \int d^3 x ~\left( \partial_i c^i\right)(x)~ \left( P_0 - \lambda \eta'\right) (x) = \lim_{r\to\infty} \int_{|\mathbf{x}|\leq r} d^3 x ~\left( \partial_i c^i\right) \left( P_0 - \lambda \eta'\right) (x) \nonumber \\
      = \lim_{r\to\infty} \left[ \int_{|\mathbf{x}|= r} d\Omega~r^2~ \left( \hat{x}_i c^i\right)~\left( P_0 - \lambda \eta'\right) (x) - \int_{|\mathbf{x}|\leq r} d^3 x ~ c^i(x)~\partial_i \left( P_0 - \lambda \eta'\right) (x)\right]. 
\end{align}
So we require that
\begin{equation}
      \label{constraint-condition-3}
      \lim_{r\to\infty}~r^2~\hat{x}_i c^i(x)=0
\end{equation}
in order that the surface term in Eq. (\ref{constraint-condition-2}) vanishes.

When $c^i$ fulfills Eq. (\ref{constraint-condition-3}), we denote the constraint in Eq. (\ref{quantum-constraint-2}) as $G(c)$, where $G$ stands for ``Gauss law":
      \begin{equation}
            G(c)=\int d^3 x ~\left( \partial_i c^i\right)~ \left( P_0 - \lambda \eta'\right) (x).
      \end{equation}

We now see that Eq. (\ref{gauss-law-in-Hamilt-1}) with Eq. (\ref{constraint-condition-1}) is the Gauss law constraint $G(B)$.

We can also consider 
\begin{equation}
      Q(D)=\int d^3x~(\partial_i D^i)~(P_0-\lambda\eta')(x),
\end{equation}
where $D^i$ may not necessarily fulfill Eq. (\ref{constraint-condition-3}). Then $Q(D)$ define the ``charges" or superselection sectors of the theory \cite{balachandran1991classical, Balachandran1992a}.

Since $G(c)|\cdot\rangle=0$, the action of $Q(D)$ on $|\cdot\rangle$ depends only on the asymptotic value of $r^2\hat{x}_iD^i$.

The constraint Eq. (\ref{constraint-1}) involves no derivative of fields and requires less careful treatment. In quantum theory the constraint it generates can be taken to be 
\begin{equation}
      G'(w)=\int d^3 x~w^iP_i,
\end{equation}
where $w^i$ are Schwartz functions.

The analogue of Eq. (\ref{quantum-constraint-2}) is
\begin{equation}
      G'(w)|\cdot\rangle = 0.
\end{equation}

The two (quantum) constraints are compatible since
\begin{equation}
      \left[ G'(w),G(c) \right] |\cdot\rangle=0.
\end{equation}

Now, there is a problem with $Q(D)$: it is not invariant under the transformations Eq. (\ref{field-transf-1}). This problem can be dealt with by imposing the quantization condition
\begin{equation}
      \label{quantization-condition-1}
      \lim_{r\to\infty} \int d\Omega~r^2 \hat{x}_i D^i (x) = n\in\mathbb{Z}
\end{equation}
and working always either with
\begin{equation}
      \label{field-transf-ind-by-charge-1}
      s(x) e^{-\frac{i}{\lambda} Q(D)} := s(x) V :=W(x)
\end{equation}
or with $V^{N_f}$.

In later considerations, we prefer to use Eq. (\ref{field-transf-ind-by-charge-1}).

Hereafter, we choose $n=1$ in Eq. (\ref{quantization-condition-1}).

\subsection{Observables}

Observables in quantum theory must commute with $G(c)$ and $G'(w)$. More generally, they must preserve the domain of the Hamiltonian. That requires their commutators with $G(c)$ and $G'(w)$ to be zero. 

Here are some examples of observables:
      \begin{itemize}
            \item[a)] Since
            \begin{align}
                  \left[ \left( P_0-\lambda\eta' \right)(x), \left(\pi-\lambda B^0 \right)(y)\right] = \left[ P_i(x),  \left(\pi-\lambda B^0 \right)(y)\right] = 0
            \end{align}
at equal times, 
\begin{equation}
      \label{obs-1}
       \pi-\lambda B^0
\end{equation}
is an observable. 

\end{itemize}

So also are
\begin{align}
({\rm b})~s(x) e^{i\eta'}, ~~~~~~ &({\rm c})~e^{iN_f \eta'}, ~~~~ ({\rm d})~ P_0, \\
    ({\rm e})~\int d^3 x~B^0,~~~ &({\rm f})~ W := s(x) V ~~ {\rm and} ~~ ({\rm g})~ V^{N_f}.
\end{align}

\subsubsection*{Remarks}

\begin{enumerate}
      \item The observable in item (a) of Eq. (\ref{obs-1}) is invariant under the substitution Eq. (\ref{field-transf-2}).
      
      \item The observable in item (e) commutes with $G(c)$,
      \begin{equation}
            \left[ \int d^3x~B^0, G(c)\right] = i\int d^3x~\partial_i c^i = 0
      \end{equation}
in view of Eq. (\ref{constraint-condition-3}). But 
\begin{equation}
      e^{iQ(D)}~\left( \int d^3x~B^0(x) \right)  ~e^{-iQ(D)} = \int d^3x~B^0(x) + n, ~~~ \textrm{ with } n\in\mathbb{Z}.
\end{equation}

      \item We should in principle smear the fields in the list (a) to (g) with appropriate test functions, say Schwartz functions. Then the algebra of observables is generated by these smeared fields\footnote{Or rather, to avoid issues of domains, by the Weyl modifications of these smeared fields as required.}.
      
\end{enumerate}

\subsection{The $\theta$-states}

Let $|\cdot\rangle_0$ denote a quantum state for $\theta=0$, or rather $\sin\theta=0$. Then the QCD $\theta$-state is
\begin{equation}
      |\cdot\rangle_{\sin\theta}=U(\theta)|\cdot\rangle_0,
\end{equation}
with
\begin{equation}
      U(\theta)=e^{i\theta\lambda \int d^3x~B^0}.
\end{equation}
This state is gauge invariant. Furthermore, the operator $U(\theta)$ commutes with $V$ if $\theta=2\pi$:
      \begin{equation}
            U(\theta)^{-1}~V~U(\theta) = V e^{-i\theta}.
      \end{equation}
That provides support to the $2\pi$-periodicity of QCD $\theta$-states.

The QCD analogue of $V$ is a winding number $1$ gauge transformation.

\subsubsection*{Remarks}

But $U(2\pi)$ does not leave the observable $P_0$ invariant and that requires explanation. That is the following.

Let us consider the vacuum expectation value $\langle \frac{P_0}{2\pi\lambda} \rangle$ of $\frac{P_0}{2\pi\lambda}$:
\begin{equation}
\langle U(\theta)^{-1}~\frac{P_0}{2\pi\lambda}~U(\theta) \rangle = \langle \frac{P_0}{2\pi\lambda} \rangle + \frac{\theta}{2\pi}
\end{equation}
Then $\frac{P_0}{2\pi\lambda}-\langle \frac{P_0}{2\pi\lambda}\rangle$ is invariant when $\frac{P_0}{2\pi\lambda}$ is conjugated by $U(\theta)$. So we can focus attention to $\langle \frac{P_0}{2\pi\lambda}\rangle$. 

Now, if we represent $U(2\pi)$ by the function $\hat{u}$ on a circle, then $\langle \frac{P_0}{2\pi\lambda}\rangle$ acts on $\hat{u}$ as its momentum by eq. (3.28). So $\hat{u},\langle \frac{P_0}{2\pi\lambda} \rangle$ describe a circle $S^1$ and its momentum.

A well-known consequence of this interpretation is that the spectrum $\sigma(\theta)$ of momentum for the $\theta$-states is $\left\{ n+\theta/2\pi: n\in\mathbb{Z}\right\}$, that is, for the vectors $U(\theta)|\cdot\rangle_0$, 
\begin{equation}
\sigma(\theta)=\left\{ n+\theta/2\pi: n\in\mathbb{Z}\right\}.
\end{equation}
Thus there is a spectral flow as $\theta$ changes by $2\pi$, with
\begin{equation}
      \sigma(\theta)=\sigma(\theta+2\pi).
\end{equation}
In this sense, the theory has a $2\pi$-period in $\theta$: $\langle \frac{P_0}{2\pi\lambda} \rangle$ is not invariant under conjugation by $U(2\pi)$, but its spectrum is.

But $U(\theta)$ shifts the vacuum expectation value of the quantum field $P_0$ and is thus spontaneously broken for generic values of $\theta$. For the circle problem, this is reflected in the fact that it shifts the domain of momentum \cite{Balachandran2012}.

\section{The Spectrum of the Hamiltonian}

We want to show that the Hamiltonian $H$ in Eq. (\ref{hamiltonian-eta-2}) describes a field with mass $\lambda R$.

First note that finiteness of energy requires that
\begin{equation}
      \int d^3x ~ \left( \partial_i \eta'\right)^2(x) < \infty
\end{equation}
which implies that 
\begin{equation}
      \partial_i \eta'(x) \to 0 ~~~ \rm{as} ~~~ |\mathbf{x}|\to \infty.
\end{equation}
Hence, it is like a $B^i$ or $c^i$ discussed previously.

For this reason we can write
\begin{equation}
      \label{condition-for-eta-field-1}
      \int d^3x~\left( \partial_i \eta'\right)^2(x)\approx \int d^3x~\left( \partial_i \eta'\right)(x)~\frac{1}{\lambda} \partial_iP_0(x).
\end{equation}

Since Eq. (\ref{hamiltonian-eta-2}) also requires that
\begin{equation}
      \int d^3x~\left( P_0-\theta\lambda\right)^2(x)<\infty,
\end{equation}
we further require that
\begin{equation}
      P_0(x)\to \theta\lambda ~~~ \rm{as} ~~~ |\mathbf{x}|\to \infty.
\end{equation}
Hence
\begin{equation}
      \partial_i P_0(x)\to 0 ~~~ \rm{as} ~~~ |\mathbf{x}|\to \infty.
\end{equation}

So we can once more substitute for $\partial_i\eta'$ in Eq. (\ref{condition-for-eta-field-1}) to find
\begin{equation}
      \int d^3x~\left( \partial_i \eta'\right)^2(x)\approx \frac{1}{\lambda^2}\int d^3x~\left( \partial_i P_0(x) \right)^2.
\end{equation}
Thus modulo constraints,
\begin{align}
      H=\int d^3 x~\left[ \frac{R^2}{2}\left(\pi-\lambda B^0 \right)^2(x)+ \frac{1}{2}\left(P_0-\theta\lambda\right)^2(x) +\frac{1}{2(\lambda R)^2}\left( \partial_i P_0\right)^2(x) \right].
\end{align}

The fields
\begin{align}
      \chi &\equiv \frac{1}{\lambda R} \left(P_0-\theta\lambda \right), \\
      \pi_\chi &\equiv R \left(\pi-\lambda B^0 \right)
\end{align}
are canonically conjugate to each other and therefore
\begin{equation}
      H=\int d^3 x~\left[\frac{1}{2} \pi_\chi^2(x) + \frac{1}{2}\left( \partial_i\chi\right)^2(x) + \frac{(\lambda R)^2}{2} \chi^2 \right] 
\end{equation}
describes a field of mass $\lambda R$.

\section{Coupling to Fermions}

The canonical chiral field for $N_f$ flavors, invariant under the transformations Eq. (\ref{field-transf-1},\ref{field-transf-2}) is the $N_f\times N_f$ matrix of fields 
\begin{equation}
      u=s~e^{i\eta'}
\end{equation}
of Eq. (\ref{u-field-1}).

We now have in addition the field
      \begin{equation}
            W=s~V
      \end{equation}
listed in section III.B.
      
The Lagrangian involving just Goldstone bosons can be written in the usual way in terms of $u$ \cite{balachandran1991classical}. Our focus here is the coupling of the Goldstone modes to fermions. In particular, we are interested on the fermion mass modification due to $W$. 

We note that $V$ is the operator implementing the winding number $1$ gauge transformations. It carries no energy and momentum\footnote{Since $V^{-1}HV=H$ and $V$ is invariant under spatial translations.} and can have a tadpole-like effect on Feynman diagrams. Since $P$ or $T$ reverses $V$, that is,
\begin{equation}
      P,T:~V\mapsto V^{-1},
\end{equation}
such tadpoles can induce electric dipole moment, or more generally flavor-diagonal $P$ and $T$ violations. Our model for electric dipole moment is based on this idea.

The left- and right-quark fields $q_L$ and $q_R$ are $N_f$-dimensional multiplets transforming by $SU(N_f)_L$ and $SU(N_f)_R$. Their standard coupling $\mu\left(\bar{q}_L u q_R+ \bar{q}_R u^\dagger q_L \right)$ to $u$ can now be generalized in a chirally invariant manner as follows:
      \begin{equation}
	    \label{mass-Lagrangian-1}
            \mathcal{L}_{q}=\mu_1\left( \bar{q}_L u q_R+ \bar{q}_R u^\dagger q_L\right) + \mu_2 \left(\bar{q}_L W q_R+ \bar{q}_R W^\dagger q_L \right).
      \end{equation}

In the $\theta$-vacuum $|u=\mathbb{1}\rangle_{\sin\theta}$ of chiral fields, $u$ has the value $1$ and $V$ has the value $e^{-i\theta}$. Thus in this vector state, $\mathcal{L}_{q}$ has the expectation value
\begin{align}
      \hat{\mathcal{L}}_q(\sin\theta) &=~_{\sin\theta}\langle u=\mathbb{1} | \mathcal{L}_{q}|u=\mathbb{1}\rangle_{\sin\theta} \nonumber \\
      &= \mu_1 \left( \bar{q}_L q_R+ \bar{q}_R q_L\right) + \mu_2 \left( e^{-i\theta}\bar{q}_L q_R+ e^{+i\theta}\bar{q}_R  q_L\right).
\end{align}

For $\theta=0$ ($\theta=\pi$), $\hat{\mathcal{L}}_q(0)$ is the standard $P$ and $T$ invariant mass term $\mu\bar{q} q$ with $\mu=\mu_1+\mu_2$ ($\mu=\mu_1-\mu_2$).

But for $\sin\theta\neq 0$, the mass term violates $P$ and $T$:
      \begin{align}
      \label{new-mass-Lagrangian-1}
      \hat{\mathcal{L}}_q(\sin\theta) &= \left(\mu_1 +\mu_2 \cos\theta\right) \bar{q} q - i\mu_2\sin\theta \left(\bar{q}_L q_R - \bar{q}_R  q_L\right).
\end{align}
We can regard $\mu_2$ as the new mass scale in the model. It cannot be determined using just the effective Lagrangian approach. Thus EDM measurements can detect only $\mu_2\sin\theta$ as (\ref{new-mass-Lagrangian-1}) shows. Presumably $\mu_2$ is of the order of the QCD scale $\Lambda_{\rm{QCD}}$. With such a value of $\mu_2$, one can also estimate $\sin\theta$ from experimental bounds on EDM. We do such estimates in a later paper \cite{Bal-edm-2012}. 

The vacuum value of $u$ is $\mathbb{1}$. Any transformation which shifts it is spontaneously broken. One such transformation is the axial $U(1)_A$ which changes $q$ to $e^{i\alpha \gamma_5}q$ and shifts $\eta'$ by $2 \alpha$. It is not a symmetry and cannot be used to eliminate the $P$ and $T$ violating term in Eq. (\ref{new-mass-Lagrangian-1}). In a future calculation, it will be explicitly shown that this term contributes to the electric dipole moment.

We can also consider mass-like terms of the baryons. For two flavors, Eq. (\ref{mass-Lagrangian-1}) is replaced by
\begin{equation}
      \mathcal{L}_N =m_1 \left(\bar{N}_L u N_R +\bar{N}_R u^\dagger N_L \right) + m_2 \left(\bar{N}_L W N_R +\bar{N}_R W^\dagger N_L  \right),
\end{equation}
where $N$ is the nucleon field. That gives
\begin{align}
      \hat{\mathcal{L}}_N(\sin\theta) &=~_{\sin\theta}\langle u=\mathbb{1} | \mathcal{L}_{N}|u=\mathbb{1}\rangle_{\sin\theta} \nonumber \\
      &= \left(m_1 +m_2 \cos\theta\right) \bar{N} N - im_2\sin\theta \left(\bar{N}_L N_R - \bar{N}_R  N_L\right).
\end{align}

The fermion propagators are thus affected by $\sin\theta$. If, in the case of quarks,
\begin{align}
      \mu(\theta) &=\mu_1 +\mu_2 \cos\theta, \\
      \mu'(\theta) &= \mu_2 \sin\theta,
\end{align}
then Eq. (\ref{new-mass-Lagrangian-1}) being $\bar{q}\left( \mu(\theta) + i\gamma_5 \mu'(\theta)\right) q$, the quark propagator for momentum $k$ is
\begin{equation}
      S(i\gamma\cdot k;\mu(\theta)) =\left(i\gamma\cdot k + \mu(\theta) + i\gamma_5 \mu'(\theta) \right)^{-1} = ~ - ~\frac{\left( i\gamma\cdot k - \mu(\theta)\right)+i\gamma_5 \mu'(\theta)}{k^2+\mu(\theta)^2+\mu'(\theta)^2}
\end{equation}
The $P$ and $T$ violating term is neatly isolated as the $\gamma_5$-term in the numerator. 

There are similar propagators for baryons.

\section{Final Remarks}

In a forthcoming paper, we will evaluate the EDM from the one-loop diagram of Fig. 1.

\begin{figure}[!ht]
  \begin{center}
      \label{fig-1}
         \includegraphics[scale=0.7]{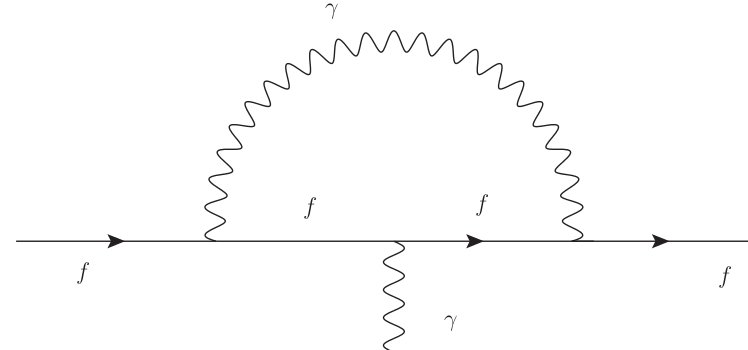}
   \end{center}
  \caption{$f$ and $\gamma$ denote fermions and photons, respectively.}
\end{figure}

One interesting aspect of such a diagram in our model is that it has no ultraviolet divergence in the parity odd sector. There is an infrared divergence in this sector, though. However it is exactly the same infrared divergence appearing in the anomalous magnetic moment computation and it is similarly canceled by soft photon emissions at the cross section level. It thus gives an unambiguous answer for EDM. This will be reported in a forthcoming paper.

\section{Acknowledgement}

The authors would like to thank Alvaro Ferraz for the hospitality at the International Institute of Physics at the Universidade Federal do Rio Grande do Norte in Natal, Brazil where discussions that led to this work were initiated. APB and ARQ also thank Prof. Alberto Ibort for the hospitality at Departamento de Matem\'aticas, Universidad Carlos III de Madrid, Spain where this work was completed. We also thank Prof. G. Marmo for discussions at the final stages of this work. APB is supported by DOE under grant number DE-FG02-85ER40231 and by the Institute of Mathematical Sciences, Chennai. ARQ is supported by CNPq under process number 307760/2009-0.


\providecommand{\href}[2]{#2}\begingroup\raggedright\endgroup

\end{document}